\documentclass[12pt]{article}%
\usepackage{physics}
\usepackage[nosort]{cite}
\usepackage{graphicx}
\usepackage{multicol}
\usepackage{amsfonts}
\usepackage{amssymb}
\usepackage{amsmath}
\usepackage{heck}
\usepackage{afterpage}
\usepackage{setspace}
\usepackage{verbatim}
\usepackage{color}
\usepackage{longtable}
\usepackage{float}
\usepackage{subcaption}
\usepackage{epsfig}
\usepackage{epstopdf}
\usepackage{adjustbox}
\usepackage{tikz}
\usepackage[margin=1in]{geometry}
\usepackage{titletoc}
\usepackage{hyperref}%
\providecommand{\U}[1]{\protect\rule{.1in}{.1in}}
\newsavebox{\mysavebox}

\hypersetup{colorlinks,citecolor=black,filecolor=black,linkcolor=black,urlcolor=black,pdftex}
\usetikzlibrary{decorations.markings}

\numberwithin{equation}{section}

\newcommand{\ba}{\begin{eqnarray}}
\newcommand{\ea}{\end{eqnarray}}

\newcommand{\cL}{\mathcal{L}}

\newcommand{\cO}{\mathcal{O}}

\newcommand{\be}{\begin{equation}}
\newcommand{\ee}{\end{equation}}
\tikzstyle{startstop} = [rectangle, rounded corners, minimum width=3cm, minimum height=1cm,text centered, draw=black, fill=blue!10]
\tikzstyle{startstop} = [rectangle, rounded corners, minimum width=3cm, minimum height=1cm,text centered, draw=black, fill=blue!10]
\tikzstyle{io} = [trapezium, trapezium left angle=70, trapezium right angle=110, minimum width=3cm, minimum height=1cm, text centered, draw=black, fill=blue!30]
\tikzstyle{process} = [rectangle, minimum width=3cm, minimum height=1cm, text centered, draw=black, fill=orange!30]
\tikzstyle{decision} = [diamond, minimum width=3cm, minimum height=1cm, text centered, draw=black, fill=green!30]
\tikzstyle{arrow} = [thick,->,>=stealth]
\begin{document}

\date{December 2024}

\title{How to Falsify String Theory at a Collider}

\institution{ASU}{\centerline{$^{1}$Department of Physics, Arizona State University, Tempe, AZ 85287, USA}}
\institution{PENN}{\centerline{$^{2}$Department of Physics and Astronomy, University of Pennsylvania, Philadelphia, PA 19104, USA}}
\institution{PENNmath}{\centerline{$^{3}$Department of Mathematics, University of Pennsylvania, Philadelphia, PA 19104, USA}}

\authors{
Matthew Baumgart\worksat{\ASU}\footnote{e-mail: \texttt{matt.baumgart@asu.edu}},
Panagiotis Christeas\worksat{\ASU}\footnote{e-mail: \texttt{pchrist8@asu.edu}},\\[4mm]
Jonathan J. Heckman\worksat{\PENN, \PENNmath}\footnote{e-mail: \texttt{jheckman@sas.upenn.edu}}, and
Rebecca J. Hicks\worksat{\PENN}\footnote{e-mail: \texttt{rjhicks@sas.upenn.edu}}
}

\abstract{The string landscape accommodates a broad range of possible effective field theories. This poses a challenge for extracting verifiable predictions as well as falsifiable signatures of string theory.
Motivated by these considerations, in this work we observe that all known stringy Standard Models support only low-dimensional representations of the gauge group. While it is in principle possible to produce contrived models with higher-dimensional representations, these generically appear in a tower of states with lighter ones in lower-dimensional representations, i.e., not in isolation. With this in mind, we consider the phenomenologically well-motivated scenario given by adding a single Majorana field in a real, $n$-dimensional representation of $SU(2)_L$ with $n \geq 5$ \textit{and nothing else}. This scenario is not realized in any known string construction, and we conjecture that this is true of string theory in general. Detection of this scenario would thus amount to falsifying the (known) string landscape. We recast existing LHC searches for new electroweak states to extract updated bounds on this class of scenarios. Improved limits from future colliders and dark matter detection experiments provide additional routes to potentially falsifying string theory.}

\maketitle

\enlargethispage{\baselineskip}

\setcounter{tocdepth}{2}

\section{Introduction}

One of the challenges in connecting string theory with observation is the vast
array of possible low energy phenomena which can in principle emerge at long distances.
For example, in the context of the string landscape (see e.g.,
\cite{Douglas:2006es, Denef:2008wq} for reviews), there is a general
expectation that many effective field theories can be accommodated by tuning
parameters, i.e., by scanning over a \textquotedblleft
discretuum\textquotedblright\ of vacua (see e.g., \cite{Bousso:2000xa,
Dine:2015xga}). Of course, since there are also expected to be a finite
(though quite large) number of such vacua, one expects there to be subtle
correlations between these effective field theories (see e.g.,
\cite{Acharya:2006zw, Heckman:2019bzm}). Another promising route is to
develop string-motivated phenomenological scenarios which would
otherwise be difficult to imagine from a purely bottom up point of view.

But making verifiable predictions is only part of having a successful theory.\footnote{Other desiderata include
the explanatory power of a framework, its satisfying a certain principle of economy, and many more criteria.}
It is also important to find concrete signatures which can \textit{falsify} a theory.

Our aim in this work will be to propose a class of phenomenological signatures which---if observed---would
immediately rule out all known string vacua, effectively falsifying string theory.
The main idea can be stated as a conjectural but very well-motivated constraint on the sorts of matter representations of the Standard Model gauge group $G_{\mathrm{SM}} = SU(3)_C \times SU(2)_L \times U(1)_Y$ which appear below the string scale:\footnote{Near and above the string scale, one expects on general grounds that all possible representations appear.}

\begin{center}
\colorbox{gray!20}{Only low-dimension representations of $G_{\mathrm{SM}}$ appear in string theory!}
\end{center}

Indeed, while it is straightforward to produce many sorts of gauge theories via string theory, the matter fields tend to come in low-dimensional representations of the gauge group. Qualitatively, this holds for perturbative open strings because the endpoints of the open string allow one to build one- and two-index tensor representations of gauge groups such as $SU(N)$, $SO(N)$, and $Sp(N)$. Away from this special case, one can indeed produce more general sorts of gauge groups and representations, but in all known string constructions the actual representations which appear are still quite low in dimension.

This motivates two natural questions:
\begin{enumerate}
    \item What is the largest dimension representation of $G_{\mathrm{SM}}$ which can appear?
    \item Can these large dimension representations appear in isolation?
\end{enumerate}
With regards to item 1), it is already challenging to produce high-dimensional representations in explicit string compactifications. For example the biggest known representation in F-theory models is the $3$-index symmetric representation of $SU(2)_L$, i.e., the $4$-plet of $SU(2)_L$.\footnote{See e.g., \cite{Ludeling:2014oba, Klevers:2017aku, Raghuram:2018hjn, Cvetic:2018xaq, Taylor:2019ots}.} Even producing the $5$-plet of $SU(2)_L$ is difficult to arrange. More broadly, there are some methods for producing higher-dimensional representations. These include, for example, free fermion constructions at higher Kac-Moody levels. Another approach is to build higher-dimension matter representations via composite operators induced by strongly coupled dynamics from an extra sector. Even so, we are not aware of a single example in the known string landscape which generates the Standard Model and also succeeds in producing the $5$-plet of $SU(2)_L$.\footnote{There are by now many constructions which contain the gauge group, chiral matter and interactions for the Standard Model. See e.g., \cite{Marchesano:2024gul} for a recent overview.}

Turning to item 2), the qualitative feature of all known methods for producing higher-dimensional representations is that these new states are actually part of an entire tower of new states, with additional lighter states in lower-dimensional representations. Said differently, the desired $n$-plet never appears in isolation. As a simple example to keep in mind, recall that in QCD-like extra sectors in which the Standard Model gauge group is viewed as a flavor symmetry, one can produce baryons in very high-dimension representations of the flavor symmetry group (viewed as containing part of the Standard Model gauge group), but there are also lighter mesons which cannot be decoupled. Decoupling all of these extra states is even more implausible.

This sets up an exciting way to potentially falsify the known string landscape, and, modulo a few well-motivated caveats, the edifice of string theory:

\begin{center}
\colorbox{gray!20}{What if an experiment detects a high-dimensional representation in isolation?}
\end{center}

There are sound phenomenological motivations for such scenarios. For example, the ``minimal dark matter'' of \cite{Cirelli:2005uq} introduces a single Majorana field in the $5$-dimensional representation of $SU(2)_L$. Getting the correct dark matter relic abundance in a thermal history for the early Universe requires this new particle to have a mass of $14$ TeV. There is an entire experimental program aimed at detecting / constraining such electroweak states, and within a decade there is good reason to believe that much of the relevant parameter space will be explored.\footnote{See e.g., \cite{Cirelli:2005uq, DelNobile:2009st, DiLuzio:2015oha, Ostdiek:2015aga, DiLuzio:2018jwd, Belyaev:2020wok,Bottaro:2021srh, Baumgart:2023pwn}. It is also worth commenting that one can entertain a higher mass $5$-plet as a dark matter candidate provided one has a non-thermal history in which a late decaying scalar dilutes the overall relic abundance.
}

More broadly, one can ask about direct constraints on high-dimensional representations, independent of their motivation as a dark matter candidate. In principle one can entertain high-dimensional representations for all of $G_{\mathrm{SM}}$, but to illustrate the main points, we focus on a single well-motivated class of possibilities in which we add a single Majorana field $\chi$ of mass $M$ in a real $n$-dimensional representation of $SU(2)_L$ \textit{and no other states}.\footnote{Similar considerations hold for related matter in high-dimensional representations. For example, string theory does not appear to accommodate scenarios involving just a scalar in a high-dimensional $n$-plet. Another canonical choice would be to take an $n$-plet of minimally coupled Dirac fields. This can be viewed as two minimal Majorana fields of the same mass. We expect similar phenomenology in all these situations. There are of course many variations involving related scenarios which would be exciting to explore in future work.}
The Lagrangian then takes the form:
\begin{equation}
\mathcal{L} = \mathcal{L}_{\mathrm{SM}} + \frac{1}{2} \overline{\chi} (i \gamma^\mu D_\mu - M) \chi.
\end{equation}

Our aim in this paper will be to extract experimental bounds on this class of scenarios using current LHC limits, essentially by recasting known limits on other searches for new electroweak states e.g., the wino. One can envision future experiments such as a proposed muon collider and other higher energy hadronic colliders which can probe even higher energy scales \cite{Cirelli:2014dsa,Han:2018wus,Capdevilla:2021fmj,Bottaro:2021snn}.  These are in addition to ongoing efforts to find a new, large electroweak multiplet as part of the direct detection \cite{Bottaro:2021snn} and indirect detection \cite{BaumgartXXX} of dark matter.

Due to electroweak symmetry breaking, higher electrically charged components of the $n$-plet turn out to be heavier, thus allowing for cascade decays which terminate with $\chi^0$, the neutral component of the $n$-plet.
The main search strategy we focus on is thus a disappearing track plus a jet (from initial state radiation) as the primary signature.

We simulate the expected results, scanning over a range of masses $M \sim 200$ GeV - $1$ TeV in steps of $25$ GeV, as well as values of $n = 3, 5,7,9$. The mass range is dictated by recasting the available wino search limits presently available from the LHC, and the range of values for the dimension $n$ is dictated by the condition that the new state does not produce appreciable (and in principle detectable) running of the $SU(2)_L$ gauge coupling in the TeV scale range.

Recasting the results of the wino search, we use current LHC data to set an updated limit on such $n$-plet scenarios.  We also project these limits to higher masses reachable by the high-luminosity LHC.

The rest of this paper is organized as follows. We begin by briefly reviewing some of the salient facts about the ``just $n$-plet scenario" in section \ref{sec:REVIEW}. In section \ref{sec:SIMULATION} we present the details of our simulation, and in section \ref{sec:RESULTS} we present our analysis results which recast the wino search limits for the present scenario. Section \ref{sec:DISC} presents some further discussion on the scenario and its implications.

The Appendices contain supplementary material which ranges from review to some original results. In Appendix \ref{app:NPLET} we provide some additional detail on some aspects of the ``just $n$-plet scenario''. Appendix \ref{app:STRING} provides a survey for building higher-dimensional representations in string theory, and thus the basis for the claim that the ``just $n$-plet scenario'' would falsify string theory. Appendix \ref{app:ATLASTRACK} reviews the ATLAS criteria for disappearing track searches.

\section{Adding Just an $n$-plet of $SU(2)_{L}$} \label{sec:REVIEW}

We shall be interested in scenarios where we extend the Standard Model by
adding a single new Majorana field $\chi$ of mass $M$ in a real representation of $SU(2)_L$ of dimension $n$. The reality (as opposed to pseudo-reality) condition requires $n$ to be odd.\footnote{In the case of pseudo-real, that is, even-dimensional representations, there are some proposals for how to realize the three-index symmetric representation of $SU(2)_L$, which would result in $n = 4$ (see e.g., \cite{Ludeling:2014oba, Cvetic:2018xaq} and also the comments in \cite{Taylor:2019ots}). Constructing stringy models with $n \geq 5$ for $n$ an integer, however, remains an open challenge. In the case of $n$ even, one can in principle arrange the hypercharge so that the lightest component of the multiplet is electrically neutral. In this case, one would still have a disappearing track phenomenology similar to the case considered here.  If $Y=0$, then even the lightest member of the multiplet would be charged, leading to different and model-dependent signatures.  Given the stakes involved in discovering such a particle, it is certainly worth future investigation.} We assume that it is neutral under $SU(3)_C \times U(1)_Y$. We are not aware of any method which produces a
``just $n$-plet scenario'' from string theory and we strongly suspect that such scenarios do not appear at all. See Appendix \ref{app:STRING} for further discussion. This state is in a spin-$j$ representation of $SU(2)_{L}$ with $2j + 1 = n$ with $n$ odd. Our theory thus contains a new set of terms (in Dirac spinor notation):
\begin{equation}
\mathcal{L}\supset c \overline{\chi}(i\gamma^{\mu}D_{\mu} - M)\chi
\end{equation}
where $c = 1$ for a Dirac field, $c = 1/2$ for a Majorana field,  $M$ is the mass,
and the covariant derivative is with respect to just the $SU(2)_{L}$ gauge bosons, i.e., $D_{\mu}=\partial_{\mu
}-igT_{j}^{a}W^{a}$, where the $T_{j}^{a}$ refer to the Lie algebra generators
for $\mathfrak{su}(2)_{L}$ in the spin-$j$ representation.

One can in principle also add direct couplings to the Higgs; this would lead to an additional degree of freedom in the model, especially with regards to the mass splittings and decay
rates for the new states. It is also more constrained from the point of view of precision electroweak constraints. To keep the discussion streamlined, we focus on the
simplest falsifiable scenario.\footnote{Of course, one can consider a wide variety of embellishments of the model. The point here is to construct a well-motivated scenario which, if observed, would falsify string theory.}

For additional details on some of the basic properties of this model, we refer to Appendix \ref{app:NPLET}. See in particular \cite{Cirelli:2005uq, Ostdiek:2015aga} for earlier work discussing collider signatures for such scenarios.

The two main parameters we can vary are $n$, the overall dimension of the representation, as well as the mass $M$. In practice, we fix a value of $n$ and then simulate different values of $M$. Jumping ahead, we find that recasting LHC searches for $3$-plets (such as the wino of the MSSM) tends to produce a limit of $M \sim O(700)$ GeV.

As for the value of $n$, the primary constraint we impose is that the running of the $SU(2)_L$ gauge coupling does not accelerate too quickly. The one-loop running
of the $SU(2)_{L}$ gauge coupling above the mass $M$ is:\footnote{Recall for $SU(N)$ gauge theory, the $k$-index fully symmetric representation (for $k>1$) has
$2$Ind$(\mathbf{R})=\frac{(N+k)!}{(k-1)!(N+1)!}$.}
\begin{equation}
\frac{d\alpha_{2}^{-1}}{dt}=\frac{b_{2}^{\mathrm{SM}}}{2\pi}-\frac{\mathrm{Ind}_{j}%
}{2\pi}=\frac{19/6}{2\pi}-\frac{1}{2\pi}\frac{(2j+2)!}{(2j-1)!3!}.
\end{equation}
where $t = \log \mu / \mu_{\mathrm{ref}}$ refers to the RG time.

Including such a state leads to an increase in the running of the gauge coupling which can in principle
drive the $SU(2)_L$ gauge theory into the non-perturbative regime. On phenomenological grounds we require
some separation between the mass $M$ and this scale of possible new physics. In practice this limits us to the following values for $n \geq 5$:\footnote{As already noted in \cite{Cirelli:2005uq}, the case $n = 5$ is rather special since the associated Landau pole is far above $M$. This is not the case for $n=7,9$, but even for $n = 9$ there is still a factor of $\sim 5$ separating $M$ and the Landau pole. For $n>9$ the theory becomes strongly coupled quite quickly anyway, and so it does not produce a ``just $n$-plet scenario'' in any meaningful sense.}
\begin{equation}
\text{Sequestered Landau Pole}: n = 5,7,9.
\end{equation}

The coupling of $\chi$ to the electroweak gauge bosons is fully fixed by having a spin-$j$ representation of $SU(2)_L$, see Appendix \ref{app:NPLET} for a brief discussion. Letting $m=j,...,-j$ denote the component of the spin-$j$ multiplet, we learn that $\chi^{m}$ has electric charge $m$. Radiative corrections from photons (and to a lesser extent $Z$-bosons) induces a mass-splitting between the component of the multiplet proportional to the charges squared; After electroweak symmetry breaking radiative corrections lead to a splitting within the $n$-plet of order $100$ MeV, with states of  higher magnitude electric charge being heavier. The lightest state is $\chi^0$, i.e., the electrically neutral one.

Let us now turn to the phenomenological signatures of this model.

\subsection{Signatures}

\begin{figure}[t!]
\begin{center}
\includegraphics[scale = 0.5, trim = {0cm 1.0cm 0cm 3.0cm}]{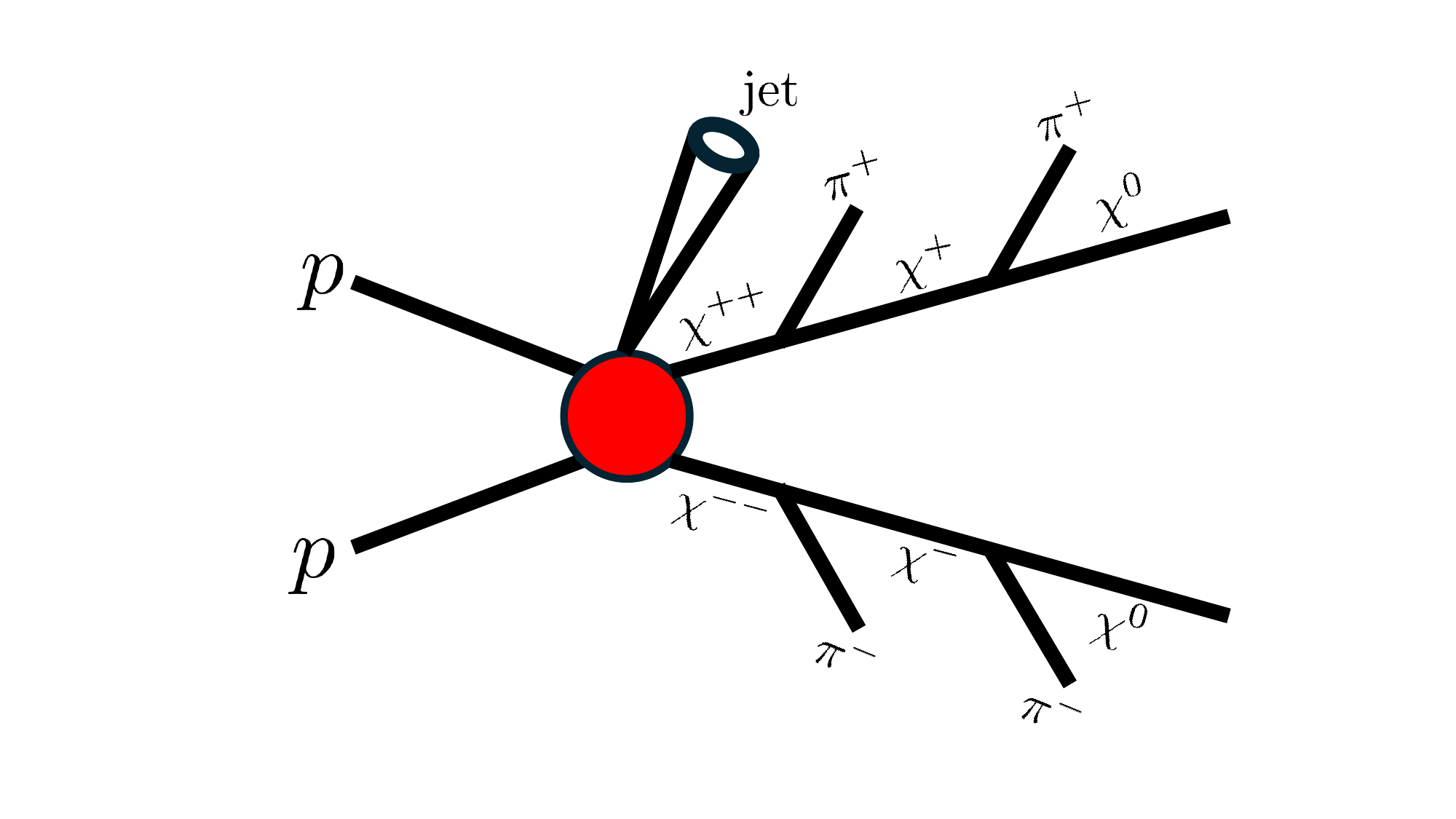}
\caption{Depiction of a production process for the $5$-plet scenario. Constituent quark / anti-quark annihilation in the protons leads to Drell-Yan production of the pair $\chi^{++}$ and $\chi^{--}$ which subsequently cascade decays to the neutral states $\chi^{0}$. Each subsequent decay emits a pion. This leads to a disappearing track + jet signature.}
\label{fig:Production}
\end{center}
\end{figure}

These electroweak multiplets share a common phenomenology with the wino of the MSSM. Existing wino searches can thus be recast to the present class of scenarios. To begin, our $\chi$ particles are color neutral, so they are produced via electroweak processes, in particular $pp \rightarrow \gamma^{\ast} / Z^{\ast} \rightarrow \chi^{m} \chi^{- m}$ and $pp \rightarrow W^{\pm \ast} \rightarrow \chi^{m} \chi^{- m \pm 1}$. There is subsequently a cascade decay (accompanied predominantly by soft charged pions) of each charged component of the multiplet to $\chi^{0}$ \cite{Cirelli:2005uq}. See figure \ref{fig:Production} for a depiction of an example process.

The $\chi^{\pm}$ state has a mean lifetime given by:
\begin{equation}
    \tau = \frac{44 \ \text{cm}}{n^2 - 1}.
\end{equation}

\begin{figure}[t!]
\begin{center}
\includegraphics[scale = 0.5, trim = {0cm 0.0cm 0cm 0.0cm}]{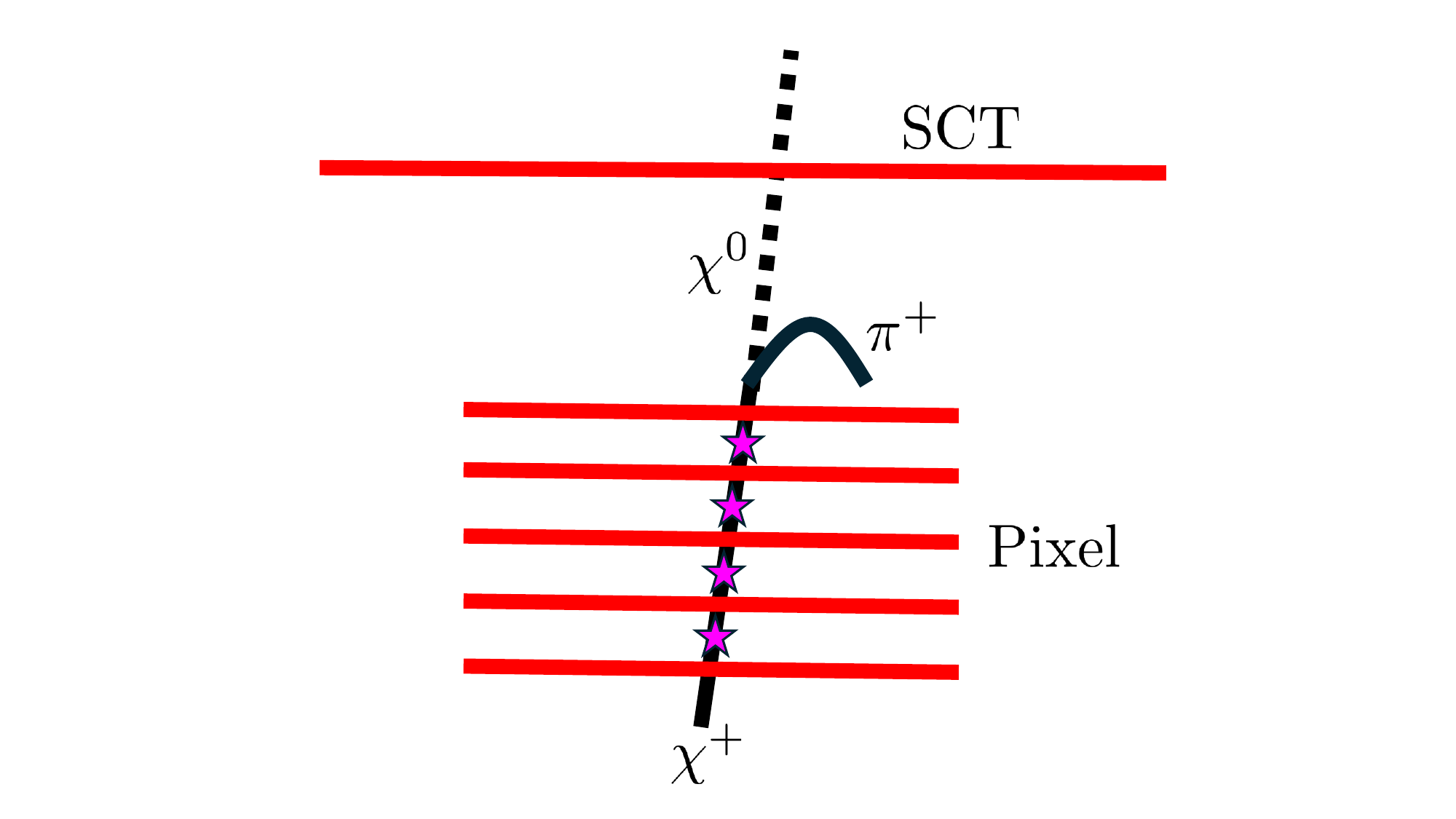}
\caption{Depiction of $\chi^{+}$ leaving a disappearing track in the Pixel (four hits indicated by the purple stars). It decays into a soft $\pi^{+}$ (circular arc) which is not seen, and to a $\chi^0$ (dashed line) which is also not detected.  SCT is the microstrip semi-conductor tracker.  The disappearing track search requires the $\chi^\pm$ to decay before reaching it.}
\label{fig:DisTrack}
\end{center}
\end{figure}

This is long enough for the $\chi^{\pm}$ state to occasionally make it through the pixel detectors of ATLAS or CMS, leaving a track that ``disappears" partway through the detector (see figure \ref{fig:DisTrack}). Disappearing tracks are not included in standard event reconstruction, so at the LHC one needs to trigger on some other kind of visible object. To correctly trigger on this scenario, we also assume the presence of at least one sufficiently hard jet from initial state radiation. This will result in missing transverse momentum that can be triggered on.\footnote{We comment that at a muon collider this jet would not be present, nor is it needed for a disappearing tracks analysis.}
The disappearing track + $E_T^{\mathrm{miss}}$ signature turns out to be the most prominent signature found in earlier studies of $n$-plet scenarios
\cite{Cirelli:2005uq, Ostdiek:2015aga, DelNobile:2009st, Han:2018wus}, and so we also adopt this as our primary signature. Disappearing track signatures are also used in long-lived chargino searches at ATLAS \cite{ATLAS:2022rme} and CMS \cite{CMS:2023mny}. Indeed, we shall recast the ATLAS wino search to set a limit on the ``just $n$-plet'' scenario.

\paragraph{ATLAS Signature}
The ATLAS long-lived chargino search looks for charginos produced either through Drell-Yan or from intermediate gluinos. The primary difference between the signal topology of these channels is that the gluino channel expects greater jet multiplicity from the gluino decay, while the electroweak channel needs at least one jet from initial state radiation (ISR). Our ``just $n$-plet'' scenario necessarily only involves the electroweak channel.  We will summarize the electroweak production mode signal region criteria and then summarize the definition of a disappearing track.

The missing transverse energy ($E_T^{\mathrm{miss}}$) trigger threshold was set between 70 GeV and 110 GeV depending on the data-taking period. Further selection cuts are applied to distinguish signal from background, and are summarized in table \ref{ATLAS_Sig_Table}.

\begin{table}[h!]
\centering
\begin{tabular}{l | l}
Signal Region Criteria & Cuts \\
\hline
Number of electrons and muons         & $0$ \\
Number of disappearing tracks       & $\geq 1$ \\
$E_T^{\mathrm{miss}}$ [GeV]              & $ > 200 $ \\
Number of jets ($p_T$ > 20 GeV) & $\geq 1$ \\
Leading jet $p_T$ [GeV]         & $> 100$ \\
$\Delta \phi^{\text{jet} - E_T^{\mathrm{miss}}}_{\text{min}}$ & $> 1.0$
\end{tabular}
\caption{ATLAS electroweak production signal region criteria \cite{ATLAS:2022rme}}
\label{ATLAS_Sig_Table}
\end{table}

Before summarizing the disappearing track criteria, some knowledge of the detector geometry is useful. The innermost detector is the pixel silicon detector (referred to as ``Pixel''). The barrel region of the Pixel consists of four layers and covers a radius of 33.3 mm to 122.5 mm and a pseudo-rapidity region of $\abs{\eta} < 1.9$.  The ATLAS search requires the chargino to traverse through the entire pixel detector, but it must decay before the microstrip semiconductor tracker (SCT) at 30 cm.\footnote{Technically, there are separate SCT detectors on each endcap of ATLAS. They are further away from the beam spot than the SCT in the barrel. There are some situations where the $\chi^{\pm}$ state can traverse through the Pixel, miss the barrel SCT but then hit one of the endcap SCTs that are further out. We take this into account in our simulation by finding the pseudo-rapidity $\eta$ of the $\chi^{\pm}$, as well as the distance $z$ along the beam-axis to determine if it will instead hit one of the endcap SCTs.}
Further out in radius are the calorimeters and muon spectrometer.

We now summarize the disappearing track cuts. Some straightforward conditions one imposes are:
\begin{itemize}
\item 4 hits in the Pixel, none in the SCT
\item Small impact parameter
\item Isolation from other objects in the event
\end{itemize}
We comment that there are also more non-trivial cuts having to do with isolating the disappearing track from calorimeter deposits. See Appendix \ref{app:ATLASTRACK} for a list of the different criteria imposed for the disappearing track search. We comment that similar criteria are applied in the analogous CMS search (see \cite{CMS:2023mny} for details), so for ease of exposition we focus primarily on just the ATLAS case.

\section{Simulation and Analysis}\label{sec:SIMULATION}

Let us now turn to the details of our simulation and analysis procedure. We utilized \texttt{FeynRules} \cite{Alloul:2013bka} to generate a model file describing the particles and their interactions. \texttt{Madgraph5} \cite{Alwall:2007st} was used for the LO matrix element calculation. The \texttt{130 NNPDF23LO} PDF was used and read by \texttt{LHAPDF6} \cite{Buckley:2014ana}. Events were showered and hadronized in \texttt{Pythia8} \cite{Bierlich:2022pfr}. Up to two additional partons were generated in \texttt{Madgraph5}, and events were matched utilizing the MLM matching procedure \cite{Mangano:2006rw,Alwall:2007fs}. Jets were clustered using \texttt{FastJet} \cite{Cacciari:2011ma} with the shower-kt algorithm.

\texttt{Delphes} \cite{deFavereau:2013fsa} was used to simulate reconstruction with the default ATLAS card. Though Delphes only propagates final-state particles through its simulation and does not simulate the inner trackers or propagate our $\chi^{\pm}$ state, it is used to simulate reconstruction for the other objects in the event.

To apply the disappearing track cuts, we use the generator-level output from \texttt{Pythia}. Additionally, ATLAS provides a \texttt{SimpleAnalysis} \cite{ATLAS:2022yru} implementation of the disappearing track search. Though we could not use this directly, we were able to use the provided efficiency maps for the disappearing track reconstruction efficiency and missing energy trigger trigger efficiency. The disappearing track efficiency map provides acceptance $\times$ efficiency for cuts we are not able to apply ourselves, such as a chi-squared-goodness of fit score. The missing transverse energy trigger efficiency map provides efficiencies for the trigger parameterized in jet energy and missing transverse energy.

Utilizing both the \texttt{Delphes} and \texttt{Pythia} output, we applied the event selection cuts and disappearing track cuts to each event. Once a final number of events was attained we scaled the final number of events to match the appropriate effective luminosity of 136 $\mathrm{fb}^{-1}$. A similar procedure can be carried out to produce limits at higher luminosity.

\section{Results}
\label{sec:RESULTS}

To obtain limits on the just $n$-plet scenarios, we recast the results of the ATLAS disappearing-track search for winos \cite{ATLAS:2022rme}.\footnote{Similar limits are set by CMS \cite{CMS:2023mny} so we focus on just recasting the ATLAS result.}
Although motivated by the presence of such a state in the MSSM, one can understand the ``electroweak channel'' minimally as testing the possibility of a new $SU(2)$-triplet fermion augmenting the Standard Model.  It is straightforward to extend the analysis to the higher-dimension representations of $SU(2)_L$ that challenge the existence of string theory.

The ATLAS search designates control and validation regions for a data-determined number of background events in their signal region.  This number, along with its uncertainty, is compared with the data in a classic ``cut-and-count'' search utilizing the $CL_s$ method to place a limit on the wino mass.  As detailed in Section \ref{sec:SIMULATION}, for our multiplets of interest we scan in mass to generate a number of events that we then subject to the ATLAS cuts and acceptance $\times$ efficiency for observing disappearing tracks.  For each model point, we thus predict a number of events that would appear in the signal region.  From this, we can perform our own $CL_s$ analysis to place an estimated limit on the BSM particle mass.

For the background-only hypothesis, having observed $n_{\text{obs}}$ events, we construct the likelihood as
\be
L_{\text{B}} = \int_{0}^{\infty} \frac{(b^\prime)^{n_{\text{obs}}} e^{-b^\prime}}{n_{\text{obs}}!} \pi_b(b^\prime,\Delta b) \, db^\prime.
\label{eq:lb}
\ee
The function $\pi_b$ is a probability distribution function (PDF) given by a truncated, normalized Gaussian which peaks at the expected number of background events in the signal region provided by ATLAS, $b$.  Its width, $\Delta b$ is the reported uncertainty on this number.  In this way, the expected number of background events $b^\prime$ in the Poisson factor is integrated over as a nuisance parameter.

Similarly, for the signal plus background hypothesis,
\be
L_{\text{S+B}} = \int_{0}^{\infty} \int_{0}^{\infty} \frac{(s^\prime + b^\prime)^{n_{\text{obs}}} e^{-(s^\prime + b^\prime)}}{n_{\text{obs}}!} \, \pi_s(s^\prime,\Delta s) \pi_b(b^\prime,\Delta b) \, ds^\prime \, db^\prime .
\label{eq:lsb}
\ee
We again convolve with a truncated-Gaussian PDF $\pi_s$, though the determination of $\Delta s$, the uncertainty on the expected number of signal events, is more involved.  There is a {\it systematic} uncertainty due to detector and analysis effects that we take from the ATLAS paper.  There is a {\it theoretical} uncertainty in the computed cross section for our model that we take from \texttt{Madgraph5}. Lastly, there is a {\it statistical} uncertainty due to the finite number of events that we generate.  To minimize this, we typically run more than 10$\times$ the number of events that would have appeared in the ATLAS data sample.  The three contributions to $\Delta s$ are added in quadrature.\footnote{The contributions to $\Delta s_{\text{syst.}}$, $\Delta s_{\text{theo.}}$, and $\Delta s_{\text{stat.}}$ are different for each production channel, and are thus also quadrature summed to give the total contributions to each for our $CL_s$ analysis.}  We summarize our signal and background parameters in table \ref{tbl:alysParams}.
\begin{table}[h!]
\centering
\begin{tabular}{|l|c|c|c|c|c|c|c|c|}
\hline
\textbf{Multiplet} & $b$ & $\Delta b$ & \textbf{$n_{\text{obs}}$} & $s$ & \textbf{$\Delta s_{\text{tot.}}$} & \textbf{$\Delta s_{\text{syst.}}$} & \textbf{$\Delta s_{\text{theo.}}$} & \textbf{$\Delta s_{\text{stat.}}$} \\
\hline
All & 3.0 & 0.7 & 3 & - & - & $0.11\, s$ & - & -\\
\hline
3 & " & " & " & 3.96 & 0.33 & 0.28 & 0.003 & 0.18 \\
\hline
5 & " & " & " & 3.48 & 0.40 & 0.18 & 0.002 & 0.38 \\
\hline
7 & " & " & " & 7.10 & 1.52 & 0.31 & 0.003 & 1.49 \\
\hline
9 & " & " & " & 4.29 & 2.28 & 0.43 & 0.003 & 2.25 \\
\hline
\end{tabular}
\caption{Parameters used in $CL_s$ analysis.  The representation-independent ones are taken from \cite{ATLAS:2022rme}.  The numbers of signal events and their uncertainties given are for the mass closest to the limit for that representation.  For the breakdown of $\Delta s$ into its various components, we take the quadrature sum of each type across production channels.}
\label{tbl:alysParams}
\end{table}

Following the $CL_s$ method\cite{ALRead_2002}, we obtain the 95\% CL limit on particle mass for each representation from the log-likelihood ratio.  Specifically, we compute
\be
\lambda(\mu) \equiv -2 \log \left( \frac{L_{\text{S+B}}(\mu)}{L_{\text{B}}} \right),
\label{eq:cls}
\ee
where $\mu$ is a factor that multiplies $s$ where it appears in $L_{\text{S+B}}$ in equation (\ref{eq:lsb}), including the $s$-dependent contributions to $\Delta s$.  We then solve for the $\mu$ value when $\lambda(\mu) = 2.71$.\footnote{The fact that the critical value of the test statistic being $2.71$ corresponds to a one-sided $95\%$ CL exclusion is nicely reviewed in \cite{Fantistats}.}$^{,}$\footnote{As an alternative to the integral convolutions in the likelihood functions (eqns.~\ref{eq:lb}) and \ref{eq:lsb}), one can instead choose the values of $b^\prime$ and $s^\prime$ that maximize log-likelihood.  In all cases we tested though, this did not change the limit.}  If $\mu > 1$, then that mass point is still viable, and conversely for $\mu < 1$.  Scanning masses in 25 GeV increments, in table \ref{tbl:recastLimits} we report the lower bound on mass for each multiplet.

There are a few qualitative comments to make on this recast analysis. First of all, we observe that in our simulation of the $3$-plet, our limit of $725$ GeV is somewhat more aggressive than that set by ATLAS ($660$ GeV, see \cite{ATLAS:2022rme}). The most likely explanation for this discrepancy is either a difference in simulating detector effects and / or the construction of the likelihood function. We find it encouraging, however, that the limit obtained from our cruder simulation efforts is within $\sim 70$ GeV of the experimentally quoted answer. Similar considerations hold for the CMS limit of $650$ GeV (see \cite{CMS:2023mny}).

The next comment concerns the overall dependence on the mass limit as a function of $n$, the dimension of the representation. Observe that the production cross section scales with $n^2$, leading to significantly more events. On the other hand, the decay rate also increases, shortening the length of the disappearing track. A second comment here is that only a tail of the momentum distribution for such decays is actually observable; at higher $n$ one is even further out on the tail of the distribution. The combined effect is a reduced limit as one increases $n$. This is not so pronounced for $n = 3,5,7$, but for $n = 9$ the limit we can set is somewhat lower ($400$ GeV).\footnote{The limit on the $9$-plet is weaker than the indirect one ($\lesssim$ 700 GeV) obtained from modifying Drell-Yan cross sections at large invariant mass due to the large modification to $\alpha_2$ \cite{Matsumoto:2017vfu}.  It is worth mentioning though, that their strongest claimed limit comes from fitting the background to a particular phenomenological background formula.  Their more conventional Monte Carlo based background method obtains a limit ($\lesssim$ 450 GeV) comparable to ours.  Should a new particle be discovered by a disappearing track search, the modification to Drell-Yan will provide a crucial test of its properties.}
\begin{table}[h!]
\centering
\begin{tabular}{|l|c|c|c|c|}
\hline
\textbf{Multiplet} & ATLAS (recast) & ATLAS \cite{ATLAS:2022rme} \\
\hline
3 & 735 & 660 \\
\hline
5 & 675 & - \\
\hline
7 & 625 & - \\
\hline
9 & 400 & - \\
\hline
\end{tabular}
\caption{95\% CL limits on the mass in GeV of a new electroweak multiplet added in isolation to the Standard Model.  For the triplet, we report the limits obtained by a recent ATLAS analysis. We then provide our results from recasting these limits.  We scan in mass in 25 GeV steps and report that lowest mass whose test statistic passes the 95\% CL.}
\label{tbl:recastLimits}
\end{table}

\begin{figure}[t!]
\begin{center}
\includegraphics[scale = 0.8]{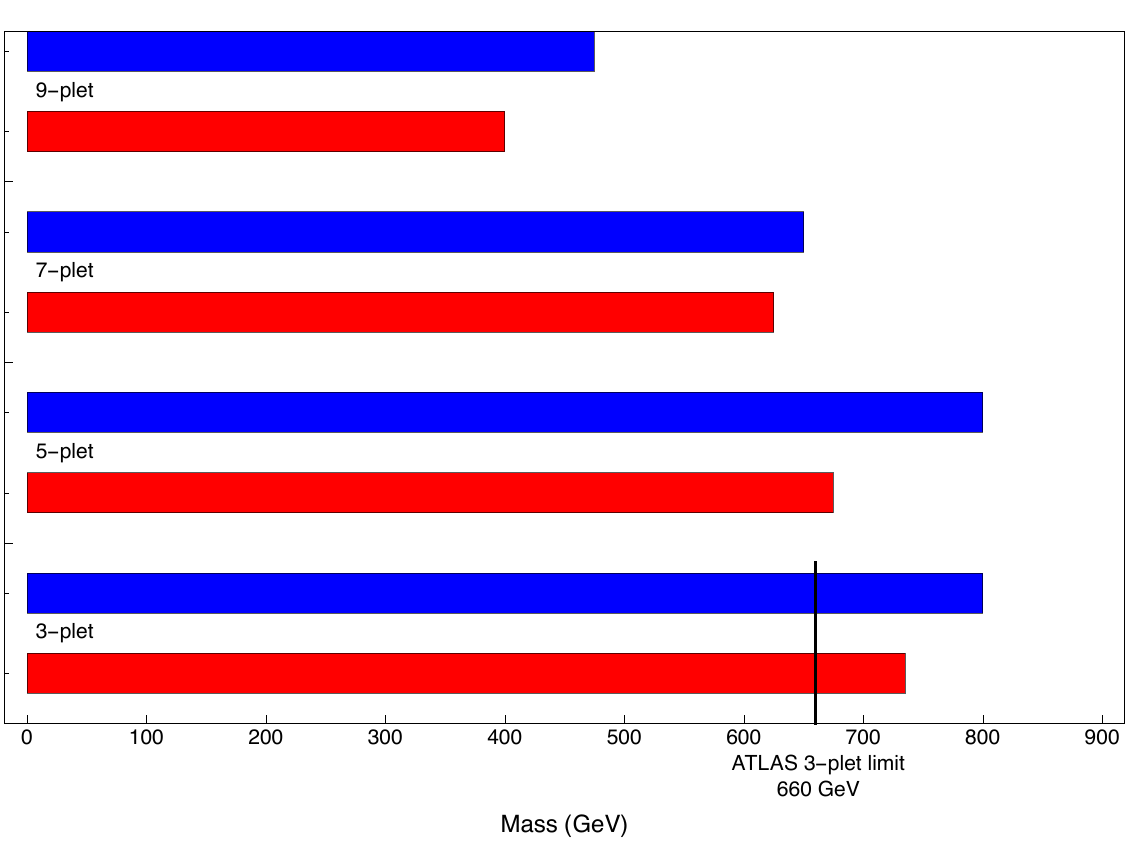}
\caption{95\% CL limits on masses for four ``just $n$-plet'' scenarios at the LHC.  The recast limits from the ATLAS disappearing track search \cite{ATLAS:2022rme} are in \textbf{red}, and our projected limits for a 3 ab$^{-1}$ run of the high-luminosity LHC are in \textbf{blue}.  The latter represent the column in table \ref{tbl:highLumi} that assumes the background scales exactly with luminosity.}
\label{fig:limits}
\end{center}
\end{figure}

Going forward, the high-luminosity LHC will make further progress on these searches.  It is beyond our scope to provide a detailed projection of the scaling of backgrounds in a cut-and-count analysis.  Instead, for the ATLAS results, we show in table \ref{tbl:highLumi} three different scenarios for 3 ab$^{-1}$ of data for how the background will scale.\footnote{We keep the systematic uncertainty, \textbf{$\Delta s_{\text{syst.}}$}, fixed at 0.11$s$ and similarly scale the background uncertainty with the background.  These values may also increase for the actual high-luminosity LHC, though the analysis is much less sensitive to increasing uncertainties than it is to the background event increases.}   If the increase in the number of background events is worse than a 50\% increase over the ratio of luminosities, then our simple cut \& count analysis will not set limits substantially stronger than the current ones.  Such a scenario would motivate innovation on the analysis side to deal with this more challenging experimental environment. For example, one could use precision Drell-Yan measurements at large invariant mass and high-luminosity to determine the running of $\alpha_2$ to improve limits on representations with $n \gtrsim 7$  beyond those given in table \ref{tbl:highLumi} \cite{Matsumoto:2017vfu,Alves:2014cda}.
\begin{table}[h!]
\centering
\begin{tabular}{|l|c|c||c|c|}
\hline
\textbf{Multiplet} & $\propto \cL$ & $\mu$ & $\propto \cL \times$1.25 & $\propto \cL \times$1.5  \\
\hline
3 & 800 & 1.47 & 800 & 800 \\
\hline
5 & 800 & 1.06 & 775 & 750 \\
\hline
7 & 650 & 1.43 & 625 & 625 \\
\hline
9 & 475 & 1.27 & 475 & 475 \\
\hline
\end{tabular}
\caption{Projected 95\% CL limits on the mass in GeV for the high-luminosity LHC at 3 ab$^{-1}$, extrapolating from the ATLAS study with 136 fb$^{-1}$.  The columns label how we scaled the number of background events relative to the increase in luminosity.  We scan in mass in 25 GeV steps and report that lowest mass whose test statistic pass the 95\% CL.  For the case where the background just scales with luminosity, we also give the $\mu$ value, which scales up the number of signal events to reach 95\% exclusion.}
\label{tbl:highLumi}
\end{table}
In figure \ref{fig:limits}, we plot the limits from both the recast analysis of \cite{ATLAS:2022rme} and our projection to the high-luminosity LHC.

\section{Discussion}
\label{sec:DISC}

In this work we have studied a class of phenomenological scenarios which is extremely
challenging, if not outright impossible to realize in string theory. The main point is that the phenomenologically well-motivated ``just an $n$-plet of $SU(2)_L$'' scenario is very hard / impossible to engineer. We do not (yet) have a no-go theorem, but there are good reasons to believe that this sort of model would pose fatal challenges for string-based model building. An additional outcome of our analysis is an updated set of current LHC limits on just $n$-plet scenarios. Let us turn to some natural extensions of the present work.

We have primarily focused on setting exclusion limits for the ``just an $n$-plet of $SU(2)_L$'' scenario. Conversely, suppose we indeed observe an excess number of events. The inverse problem of determining the $n$-plet mass $M$, as well as the specific value of $n$ would clearly be an important task, since $n \geq 5$ does not arise in string theory.\footnote{Discovering a state with $n=3$, like the chargino of SUSY or a Kaluza-Klein excitation of a $W$ boson would be a watershed moment for particle physics, but $SU(2)_L$ triplets do not pose a challenge to string theory.  In fact, the existence of $SU(2)_L$ gauginos is essentially a prediction of the superstring.} The measurable trio of particle mass, decay length (see equation (\ref{eq:decayrate})), and production rate would strongly constrain the possible value of $n$.  Other signatures / search strategies can then be used to distinguish between these possibilities. It would be exciting to investigate this question further.

On the formal side, it would of course be desirable to have a complete no-go theorem that the ``just $n$-plet'' scenario cannot be realized in string theory. While this would appear to require a deep knowledge of non-perturbative quantum gravity, all indications point to this scenario being ruled out. As a more concrete near-term goal, one might hope to prove that no perturbative string theory can realize such a scenario. Complementary to this, sharpening constraints on the mass spectrum of strongly coupled quantum field theories would also rule out scenarios in which the $n$-plet emerges as a composite state.

On the observational side, the next decade looks to be especially promising for
testing the possibility that a simple $SU(2)_L$ $n$-plet is the lowest-lying state beyond the weak scale. There are updated direct and indirect dark matter detection experiments on the horizon, and future colliders such as a proposed muon collider or high energy hadronic facility would all provide improved bounds and limits on this class of scenarios.

A further comment is that here we have focused on a specific, quite minimal extension of the Standard Model. One can envision widening our scope to bosonic states with the same representation content, as well as far more general states in high-dimensional representations of the Standard Model gauge group. One can also envision more intricate interactions which lead to distinctive signatures. Again, the key point is that a positive detection for \textit{any} such scenario would amount to ruling out the entire (known) string landscape.

More broadly speaking, we believe it is important to develop phenomenological scenarios which---\textit{if detected}---would falsify the (known) string landscape.\footnote{Depending on one's taste (or lack thereof), one might include evidence for or against cosmic inflation in this category. Other more drastic signatures include searches for breakdowns in unitarity itself \cite{Distler:2006if}.} Indeed, the value of a theory lies not just in its predictions, but also in what it cannot accommodate.\footnote{It is of course worth noting that the preponderance of evidence points to string theory being a valuable and correct framework for addressing many questions concerning quantum gravity; consistency of effective field theories coupled to gravity; motivating new scenarios for physics beyond the Standard Models of particle physics and cosmology; and as a general purpose tool in constructing and studying quantum field theories, especially at strong coupling.}

\newpage

\section*{Acknowledgements}

We thank O. Aharony, K. Dienes, T. Dumitrescu,
L. Gutierrez Zagazeta, M. Hank, F. Hassler, K. Intriligator, Z. Komargodski, A. Maloney, B. McGovern, P. Meade, I.V. Melnikov, B. Ostdiek, D. Rankin, B. Rosser,
E. Thomson, A.P. Turner, and C. Vafa for helpful correspondence and discussions. We also thank O. Aharony, M. Hank, I.V. Melnikov, and A.P. Turner for comments on an earlier draft. MB thanks the UPenn theory group for hospitality during part of this work.
JJH thanks the Harvard Swampland Initiative for hospitality during part of this work.
MB and PC are supported by the DOE (HEP) Award DE-SC0019470.
The work of JJH is supported by DOE (HEP) Award DE-SC0013528,
and BSF grant 2022100. RH is supported by an NSF Graduate Research Fellowship.


\appendix


\section{Electroweak $n$-plet of $SU(2)_L$} \label{app:NPLET}

In this Appendix we briefly review some aspects of an $n$-plet of $SU(2)_L$. We consider both the case of a
Dirac $(c = 1)$ and Majorana $(c = 1/2)$ particle in a real $n$-dimensional representation.
To set notation, we denote by $T^a$ the Lie algebra generators for $\mathfrak{su}(2)_L$, i.e., $[T^{a}%
,T^{b}]=i\varepsilon^{abc}T^{c}$. We also introduce the standard raising and
lowering operators for $\mathfrak{su}(2)_{L}$, i.e. $T^{\pm}=T^{1}\pm iT^{2}$.

By inspection, we observe that the electroweak bosons interact with the $\chi$
particles:
\begin{align}
\mathcal{L}  &  \supset c\frac{g}{\sqrt{2}}\overline{\chi}W_{\mu}^{+}\gamma
^{\mu}T^{+}\chi= c\sqrt{j(j+1)-m(m+1)}\left(  \overline{\chi}_{m+1}\gamma^{\mu
}\chi_{m}\right)  \frac{gW_{\mu}^{+}}{\sqrt{2}}\\
\mathcal{L}  &  \supset c\frac{g}{\sqrt{2}}\overline{\chi}W_{\mu}^{-}\gamma
^{\mu}T^{-}\chi= c\sqrt{j(j+1)-m(m-1)}\left(  \overline{\chi}_{m-1}\gamma^{\mu
}\chi_{m}\right)  \frac{gW_{\mu}^{-}}{\sqrt{2}}\\
\mathcal{L}  &  \supset c g\overline{\chi}A_{\mu}^{3}\gamma^{\mu}T^{3}%
\chi= cm\left(  \overline{\chi}_{m}\gamma^{\mu}\chi_{m}\right)  \left(  \frac
{g}{\cos\theta_{W}}Z_{\mu}^{0}+eA_{\mu}\right)  ,
\end{align}
where in the above we used the explicit matrix representatives for the $\mathfrak{su}(2)_L$ generators, $\cos\theta_{W}=g/\sqrt{g^{2}+g^{\prime2}}$, and
$e=g\sin\theta_{W}$.

Much of the phenomenology of the scenario is dictated by the mass splitting
within the $n$-plet, as well as the resulting decay rate to each subsequent
member of the multiplet. Observe that in the absence of electroweak symmetry
breaking, all components of the multiplet have the same mass. After electroweak symmetry breaking,
radiative corrections from the vector bosons generate a mass
splitting between the different components. In \cite{Cirelli:2005uq} the mass
splittings were studied in general. In particular, in the limit where
$M/M_{W,Z}\gg1$, the mass splitting between different electric charge
components of the multiplet have a simple form:%
\begin{equation}
M_{Q}-M_{Q^{\prime}}\simeq\left(  Q^{2}-Q^{^{\prime}2}\right)  \frac
{\alpha_{2}s_{W}^{2}}{2}M_{W}\simeq\left(  Q^{2}-Q^{^{\prime}2}\right)
\times166\text{ MeV.}%
\end{equation}
The bottom component of the multiplet is stable against further decay (there
is a $\mathbb{Z}_{2}$ symmetry which protects it),\footnote{In general this symmetry can be broken by higher-dimension operators, which has important implications for the $\chi^0$ state's cosmological stability.  One of the primary motivations for considering the $\mathbf{5}$-plet as a dark matter candidate is that its decay only occurs at dimension-6 ($\cO \sim \chi L HHH$). However, these irrelevant operators do not affect the $\chi^0$'s collider phenomenology.} but the higher components of the multiplet
are expected to eventually cascade decay down to this lightest state. On the
timescale of collider signatures, the charge $\vert Q \vert >1$ states decay within the
inner tracker. However, the final charged state decays to the neutral
component with the following rates and branching ratios \cite{Cirelli:2005uq}:%
\begin{align}
\Gamma(\chi^{+} &  \rightarrow\pi^{+}\chi^{0})=\left(  n^{2}-1\right)
\frac{G_{F}^{2}V_{ud}^2\left(  \Delta M\right)  ^{3}f_{\pi}^{2}}{4\pi}%
\sqrt{1-\left(  \frac{M_{\pi}}{\Delta M}\right)  ^{2}}\text{ \ }(\text{BR}%
\sim97.7\%) \label{eq:decayrate}\\
\Gamma(\chi^{+} &  \rightarrow e^{+}\nu_{e}\chi^{0})=\left(  n^{2}-1\right)
\frac{G_{F}^{2}\left(  \Delta M\right)  ^{5}}{60\pi^{3}}\text{ \ }%
(\text{BR}\sim2.05\%)\\
\Gamma(\chi^{+} &  \rightarrow\mu^{+}\nu_{e}\chi^{0})=0.12\times\Gamma
(\chi^{+}\rightarrow e^{+}\nu_{e}\chi^{0})\text{ \ }(\text{BR}\sim0.25\%).
\end{align}
For example, this leads to a rest-frame lifetime of the $\chi^+$ from the quintuplet of $\approx 0.06$ ns, which can easily give a macroscopic decay length.

\section{Stringy High-Dimensional Representations} \label{app:STRING}

In this Appendix we provide some additional details on why high-dimensional
representations are difficult---though not impossible---to realize in string theory.
Additionally, we motivate the conjecture that the ``just $n$-plet of $SU(2)_L$'' scenario cannot be realized. We focus on the case where the new state is a fermion, but comment that quite similar considerations hold for new bosonic states as well.
Indeed, in supersymmetric models one often has both sorts of states anyway.

There are two constraints we need to impose right from the start.
First, we need to have a gauge group which can accommodate the standard electroweak doublets, so a gauge group
such as $SO(3)$ will not work. Second, we need to find matter in the $n$-plet of $SU(2)_L$ for some $n \geq 5$. As throughout, we take $n$ to be odd (i.e., real rather than pseudo-real representations) so there is no constraint from anomalies. There are two well-substantiated conjectures that we put forward:
\begin{itemize}
\item No (known) stringy Standard Model has $n \geq 5$ matter
\item Adding just the $n$-plet for $n \geq 5$ is not possible in string theory
\end{itemize}
In both cases, our discussion falls short of a no-go theorem,
but the state of the art suggests that both claims are quite plausible.

To begin, we review some of the canonical approaches to
realizing matter representations via perturbative open and closed string
theories, as well as their extensions to strongly coupled regimes.

After this, to illustrate some of the relevant issues, we also discuss how one can use strongly coupled dynamics to build bound states with high-dimensional representations. Such scenarios can be implemented in many stringy setups, but fitting this with various qualitative phenomenological constraints is far more challenging. Additionally, in all the examples we encounter, one finds not just the $n$-plet of $SU(2)_L$ but also
lower-dimensional representations such as the triplet which are lighter or nearby in mass. Motivated by these considerations, we then give a more precise sense in which models such as the \textquotedblleft just $n$-plet scenario\textquotedblright\ for $n \geq 5$ do not arise in string theory.

\subsection{Matter with Open Strings}

\begin{figure}[t!]
\begin{center}
\includegraphics[scale = 0.5, trim = {0cm 5.0cm 0cm 3.0cm}]{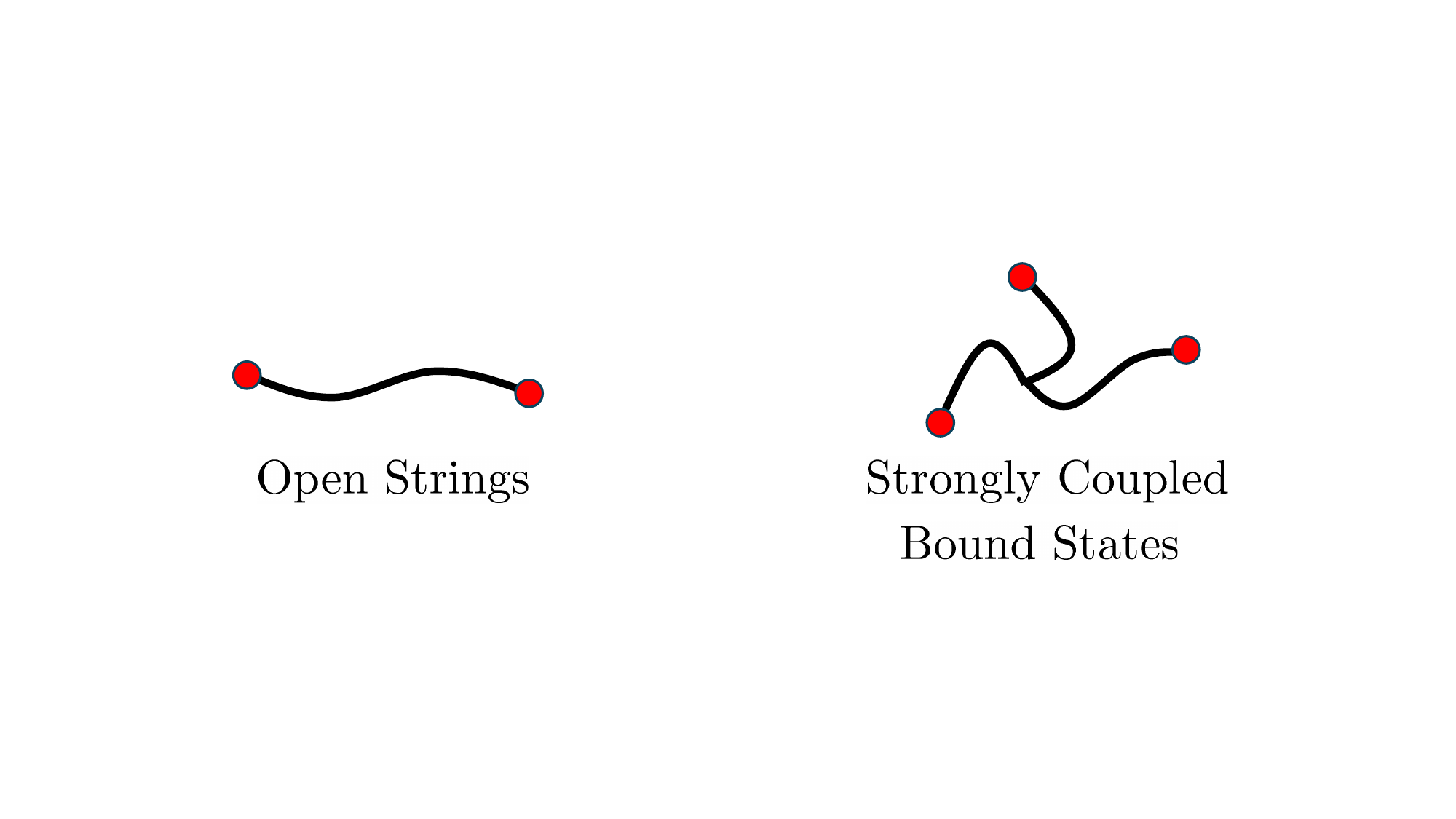}
\caption{Depiction of a perturbative open string, as well as a strongly coupled bound state of open strings.
The available representations which can be built with perturbative open strings are limited
to one- and two-index representations; the indices can be associated with the endpoint of the strings.
At strong coupling one can in principle form additional bound states associated with multi-prong string
configurations. Even so, the range of options is quite limited.}
\label{fig:OpenStringMultiProng}
\end{center}
\end{figure}

To begin, we start by discussing how stringy Standard Models arise via open
strings. We begin by working with perturbative open strings and then extend
our discussion to the case of strongly coupled bound states. Recall that in a
perturbative open string, there are two endpoints and these can be decorated
by an additional Chan-Paton factor. These endpoints should be viewed as
terminating on a Dirichlet brane (D-brane). Each such Chan-Paton factor
indicates a single fundamental index of a corresponding gauge
group.\footnote{For a helpful introduction to Chan-Paton factors, see e.g.,
the books \cite{Polchinski:1998rq, Polchinski:1998rr, Johnson:2023onr, Becker:2006dvp}.} For example, an open string which stretches from a single stack back to itself can
support gauge fields in the adjoint representation. Allowing for orientation
reversal of open strings, this general construction allows one to realize all
of the classical Lie algebras,\footnote{Since we are mainly concerned with
possible representations of a given Lie group / Lie algebra we shall not dwell
on the global form of the Lie group in question and will freely interchange
Lie group and Lie algebra conventions (by abuse of notation).} namely $SU(N)$,
$SO(N)$ and $Sp(N)$ (as well as $U(1)$ factors). Since open strings carry two
endpoints, we also see that the sorts of matter fields (spin $0$ and spin
$1/2)$ which can be accommodated in this approach consist of two-index
representations of a single gauge group. For example, for $SU(N)$ this
consists of the adjoint representation as well as the two-index symmetric and
anti-symmetric representations. Similar considerations hold for the other
classical Lie algebras. One can also achieve matter in a single index
representation by considering bifundamental representations for products of two simple gauge groups $G_1 \times G_2$. We interpret this as an open string
which stretches between two stacks of D-branes. Note in particular that matter
in the $\mathbf{5}$ of $SU(2)$ would have required a 4-index symmetric representation.
The essential point we draw from these considerations is that at least at the
level of \textit{perturbative} open strings, the sorts of matter
representations which can actually be realized are rather limited; we have
matter fields in either a single index or two index representation. See figure \ref{fig:OpenStringMultiProng} for a depiction.

Of course, there are well-known limitations to working with perturbative open
strings. For example, even the possible gauge groups which can be realized in
this way excludes all of the exceptional Lie algebras as well as important
ingredients such as the spinor representation of $SO(N)$. In type IIB\ string
theory these more general possibilities can all be realized by working in
terms of non-perturbative bound states of strings known as string junctions.
One then considers not just the fundamental string (F1-string), but also its
S-dual the D1-brane (D1-string) and dyonic bound
states of these strings known as $(p,q)$ strings. For our present
considerations, the essential point is that such bound states lead to string
junctions with more than two endpoints. As such, this provides a promising
route to realizing representations with more indices. References
\cite{Gaberdiel:1997ud, Gaberdiel:1998mv} explicitly show how to realize the
exceptional Lie algebras and some representations in terms of such string
junctions (see also \cite{Hassler:2019eso}).

To illustrate, consider matter in the $\mathbf{78}$ of $E_{6}$, namely the
adjoint representation of $E_{6}$. One can now realize representations such as
the $\mathbf{20}$ of $SU(6)$, i.e., the three-index anti-symmetric of $SU(6)$
via the decomposition:%
\begin{align}
E_{6} &  \supset SU(6)\times SU(2)\\
\mathbf{78} &  \rightarrow(\mathbf{35},\mathbf{1})+(\mathbf{1},\mathbf{3}%
)+(\mathbf{20},\mathbf{2}).
\end{align}

Such breaking patterns are also straightforward to realize in M-theory and
F-theory backgrounds.\footnote{Recall that M-theory on a small circle reduces
to perturbative IIA\ string theory \cite{Hull:1994ys, Witten:1995ex}. F-theory
is a geometrized characterization of type IIB\ string theory in which the
position dependent profile of the axio-dilaton is geometrized as the shape
modulus of a complex two-torus \cite{Vafa:1996xn, Morrison:1996na,
Morrison:1996pp}.} In this setting, one begins with a higher-dimensional gauge
theory which also includes adjoint valued matter. Working with backgrounds
where the adjoint valued matter and the gauge fluxes are switched on then
produces breaking patterns which can reproduce many of the successful breaking
patterns used in purely 4D\ GUTs (often with fewer model building
complications). See e.g., \cite{Acharya:2001gy, Beasley:2008dc, Beasley:2008kw, Donagi:2008ca, Donagi:2008kj} as well as the reviews \cite{Acharya:2004qe, Heckman:2010bq, Weigand:2018rez}.

One of the subtle points in this method is that accommodating the gauge group
of the Standard Model typically imposes rather strong restrictions on
realistic breaking patterns. For example, one of the best-motivated routes to
realizing the Standard Model factors through an intermediate GUT\ group. To
illustrate, we start with the $\mathbf{248}$ of $E_{8}$, i.e., the adjoint
representation of $E_{8}$ and consider the decomposition first to
$SU(5)_{\text{GUT}}\times SU(5)_{\bot}$:%
\begin{align}
E_{8} &  \supset SU(5)_{\text{GUT}}\times SU(5)_{\bot}\label{eq:E8decomp}\\
\mathbf{248} &  \rightarrow(\mathbf{24},\mathbf{1})+(\mathbf{1},\mathbf{24}%
)+(\mathbf{5},\overline{\mathbf{10}})+(\overline{\mathbf{5}},\mathbf{10}%
)+(\mathbf{10},\mathbf{5})+(\overline{\mathbf{10}},\overline{\mathbf{5}}),\label{eq:E8decomp2}
\end{align}
where the $\mathbf{5}$ denotes the fundamental representation (single
index)\ and the $\mathbf{10}$ denotes the two-index anti-symmetric
representation. As is well known, this suffices to reach the representation
content present in the Standard Model, but it does not support more exotic
representations such as the $\mathbf{5}$ of $SU(2)_{L}$.

To date, the vast majority of realizations of the Standard Model within string
theory actually proceed via such low-dimensional representations, i.e., either
through perturbative open string constructions or via a GUT\ group breaking
pattern as in line (\ref{eq:E8decomp}) - (\ref{eq:E8decomp2}). In principle one could contemplate a more exotic embedding of the Standard Model gauge group in $E_{8}$. For
example, one could consider the diagonal embedding of $SU(5)_{\text{diag}%
}\subset SU(5)_{\text{GUT}}\times SU(5)_{\bot}\subset E_{8}$, but even this would not produce a four-index symmetric representation.

Let us also comment that in more exotic F-theory backgrounds with additional
tunings / non-geometric ingredients it is sometimes possible to engineer more
exotic representations (see e.g., \cite{Ludeling:2014oba, Klevers:2017aku, Cvetic:2018xaq} as well as the comments in \cite{Taylor:2019ots}), but even so, engineering the $\mathbf{5}$-plet as part of a Standard Model gauge group has (to date)\ not been realized.\footnote{See \cite{RaghuramTurner:2025XXX} for recent progress on realizing the $\mathbf{5}$ of $SO(3)=SU(2)/%
\mathbb{Z}
_{2}$. Here, it is important to note that the global form of the gauge group
excludes matter in the doublet representation of $SU(2)_L$. It is also worth noting that the two-index symmetric traceless representation of $SO(N)$ is, for $N = 3$ just the $\mathbf{5}$-plet of $SO(3)$.}

The general point we make here is that in the context of building stringy
Standard Model-like vacua with high-dimensional representations, engineering
the $\mathbf{5}$-plet of $SU(2)_L$ appears challenging / impossible via this route.

\subsection{Matter with Closed Strings}

Historically the first examples of string backgrounds which include the
Standard Model gauge group were generated via perturbative closed strings,
i.e., the $E_{8}\times E_{8}$ heterotic string \cite{Gross:1984dd,
Gross:1985fr, Gross:1985rr, Candelas:1985en}. In this setting, the vast
majority of semi-realistic backgrounds arise from a 2D worldsheet CFT\ in
which the $E_{8}$ is generated via a 2D current algebra. In the target space
this appears as a 10D\ super Yang-Mills theory sector, which is subsequently
compactified to produce the relevant gauge groups and matter content of the
Standard Model. In this case, the same representation theory based
restrictions discussed above still apply, leading to rather strong
restrictions on the available representation content for matter charged under
the Standard Model gauge group.

Broader possibilities are available, however, since one can in principle
consider a more general 2D\ CFT current algebra. Here, the main idea is that
global symmetries of the 2D worldsheet CFT inevitably become \textit{gauge} symmetries in the target
space. In particular, we are interested in models where the worldsheet has an $SU(2)_L$ global symmetry.

A particularly well-studied
example involves higher Kac-Moody levels, as in references
\cite{Aldazabal:1994zk, Dienes:1995sq, Dienes:1996yh, Dienes:1996du}. The key
feature of these constructions is that one can in principle realize low
dimension conformal primary operators in high-dimensional representations. As
such, these states can in principle be part of the massless spectrum of a
string background.\footnote{Recall that in a CFT\ with left- and right-moving
conformal weights $(h_{L},h_{R})$, the massless sector of the string theory
requires weights $h_{L}=h_{R}=1$. Massive string states correspond to taking
$h>1$. These appear at the string scale and lead to a tower of states anyway
(not just a single isolated state). Tachyonic states appear for $h<1$. In a
given worldsheet CFT\ one often has different CFT\ operators so even if the
weight from one sector is less than one, the total conformal weight can still
be one. In this case, the total weight for a composite operator needs to add
up to unity to make a massless state.}

To illustrate, consider a putative chiral $SU(2)$ current algebra at level
$k\in%
\mathbb{Z}
_{>0}$. The central charge of this CFT sector is:
\begin{equation}
c=\frac{3k}{k+2}\leq3,
\end{equation}
where the upper bound is only saturated at $k\rightarrow\infty$ (a classical
background). A primary operator in a spin$-j$ representation of $SU(2)$ has
weight:%
\begin{equation}
h_{j}=\frac{j(j+1)}{k+2},\label{eq:hj}%
\end{equation}
and so in principle we can have very large values of $j$ with low weight. It
is worth noting here that this sort of construction only provides a set of
\textit{candidate} massless states in a given string spectrum. Indeed, to get
a massless state we need the total conformal weight to be exactly one. One can
accomplish this by assuming that the worldsheet CFT\ contains another sector
so that the actual operator of interest is of the (schematic) form:%
\begin{equation}
\mathcal{O}_{j,\text{full}}=\mathcal{O}_{j}\mathcal{O}_{\text{extra}%
},\label{eq:OjOextra}%
\end{equation}
so that the total weight $h_{j}+h_{\text{extra}}=1$. If one cannot find such
an extra sector and a suitable operator such that the total weight can be
raised to one, then one either has a tachyon in the spectrum or one must
include a suitable GSO\ projection to remove the offending state from the low
energy spectrum.

Group theoretically, working with Kac-Moody levels leads to non-trivial
embeddings of the Standard Model\ gauge group in a bigger product gauge group
structure \cite{Dienes:1995sq, Dienes:1996yh, Dienes:1996du}. We are unaware
of an explicit example which contains the Standard Model as well as the
$\mathbf{5}$ of $SU(2)_{L}$, but it is plausible that this, or related
construction methods based on non-geometric backgrounds (see e.g.,
asymmetric orbifolds \cite{Narain:1986qm, Baykara:2024vss})\ can yield such states.

That being said, these sorts of constructions appear to always come with
matter in a wide variety of representations, including the $\mathbf{3}$ of
$SU(2)_{L}$. To see the issue, let us return to equation (\ref{eq:hj}).
Suppose we have managed to build a model with a massless spin-$j$
representation. Consider next a spin $l<j$ representation. Observe that this
has lower weight:
\begin{equation}
l<j\Rightarrow h_{l}<h_{j}.
\end{equation}
Returning to our discussion near (\ref{eq:OjOextra}), we now face the issue of
how to remove the $\mathcal{O}_{l}$ operator from the string spectrum.\footnote{A
superficial interpretation of the equations presented might suggest engineering a
model where we get the $\mathbf{5}$-plet from setting $j = 2$ and $k = 4$ so that $h_{j = 2} = 1$,
producing a massless state. The issue with this approach is what to do with all the \textit{additional}
$h_{l} < 1$ states with $l < 2$. At a more detailed level, increasing
the level so much also tends to produce further difficulties with the values of the gauge couplings (see e.g., \cite{Dienes:1996du}).}
It is not altogether clear whether this can be accomplished since the condition of
modular invariance generally requires one to not \textquotedblleft thin
out\textquotedblright\ the spectrum of chiral primary operators. There are
some known exceptions to these considerations based on maverick coset
constructions\footnote{We thank I.V. Melnikov for helpful correspondence on
this point.} (see e.g.,\cite{Frohlich:2003hg}) but as far as we are aware
there is no known way to get \textquotedblleft just the $\mathbf{5}$-plet and
nothing else.\textquotedblright\footnote{Indeed, it is unclear to us how one can keep the spin$-j$ state massless and remove the spin$-l$ state, all whilst retaining both a tachyon free and modular invariant partition function. We thank K. Dienes, F. Hassler, I.V. Melnikov, and C. Vafa for helpful comments.}

\subsection{Building Exotic Representations via Strong Coupling}

Having reviewed some variants on open and closed string constructions of
representations in Standard Model-like vacua, we now turn to a general method
for realizing exotic representations which can be accommodated in explicit string backgrounds.
The main feature we wish to stress is that even though we can indeed realize these
higher-dimension representations, the resulting scenarios always contain
other, lighter states in smaller representations. This is different from the
\textquotedblleft just $n$-plet\textquotedblright\ type scenario we
have investigated in this work.

\begin{figure}[t!]
\begin{center}
\includegraphics[scale = 0.5, trim = {0cm 5.0cm 0cm 3.0cm}]{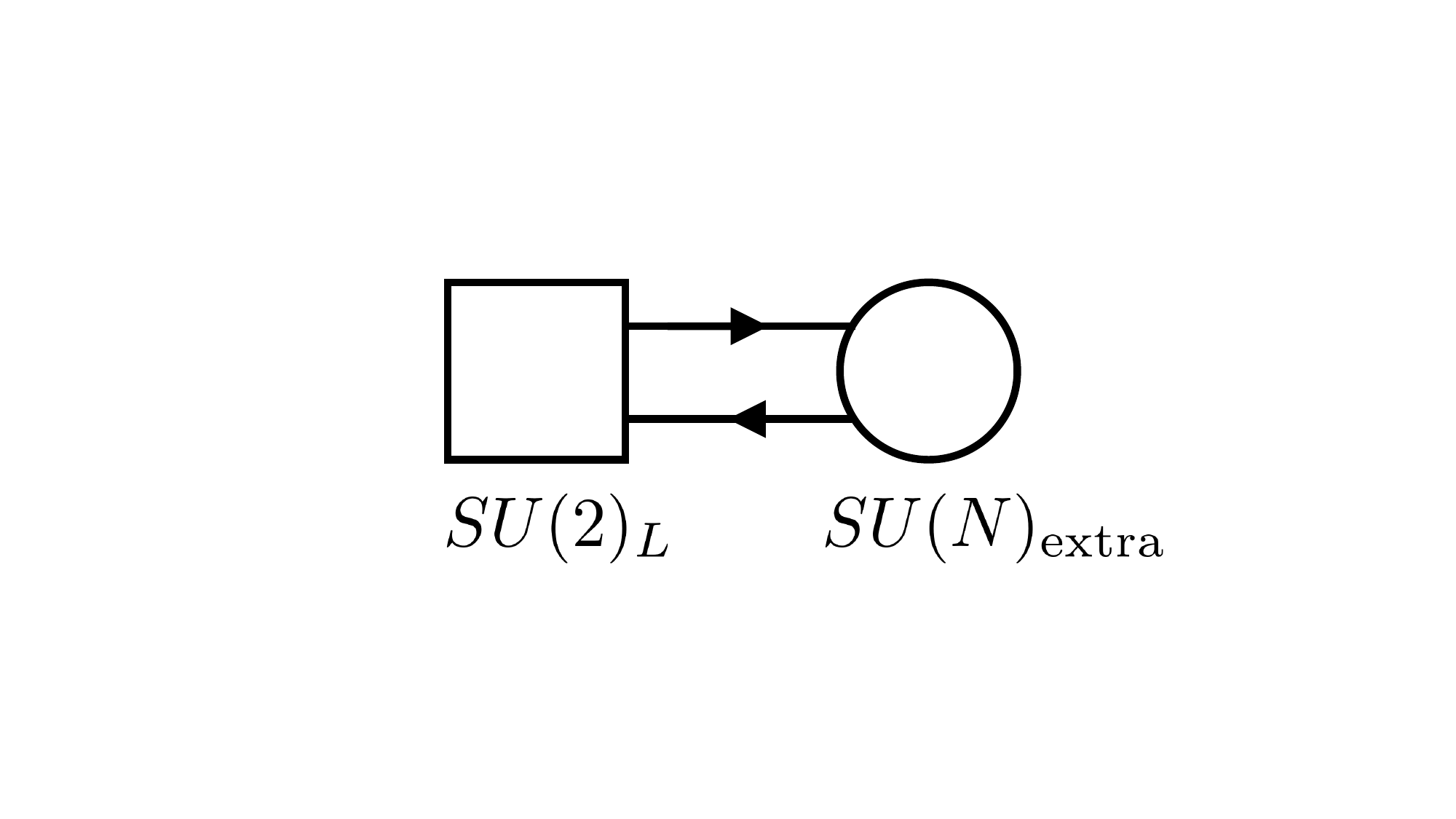}
\caption{Depiction of quiver / moose diagram where there is a strongly coupled gauge theory
with gauge group $SU(N)_{\mathrm{extra}}$ (circle), and in which the $SU(2)_L$ factor of the Standard Model gauge group appears as a weakly gauged flavor symmetry (square). The oriented lines connected the two groups are messenger fields in representations of both gauge groups. Strong coupling dynamics in the extra sector can lead to bound states in bigger representations of $SU(2)_L$, but generically these are accompanied by towers of states which cannot be decoupled. One can view the connecting lines as open strings which end on different branes used to build the respective gauge groups.}
\label{fig:ExtraSector}
\end{center}
\end{figure}

The main idea can be stated in bottom up terms: We wish to view the $n$-plet as a
composite operator which is really a strongly coupled bound state comprised of
``preons'' built from lower-dimensional representations. To build explicit
examples of high-dimensional representations we consider a generic
extra sector in which the SM\ gauge group appears as a weakly gauged flavor
symmetry. See figure \ref{fig:ExtraSector} for a representative example.

As a warmup, consider $SU(N)$ gauge theory with $N_f$ quark flavors. We split the quarks into Weyl fermions and consider left-handed fermionic fields $\psi_{(s,f)}^{g}$ where the label $s$ indicates the spacetime spin, $f$ is an index for the anti-fundamental of $SU(N_f)$ and $g$ is an index for the fundamental of $SU(N)$. We denote by $\widetilde{\psi}$ the left-handed Weyl fermions in the conjugate representations of the flavor and gauge groups. In this example, there is a gauge invariant baryonic operator:
\begin{equation}
\mathcal{B} = \varepsilon_{g_1...g_N} \psi_{(s_1,f_1)}^{g_1} ... \psi_{(s_N,f_N)}^{g_N}.
\end{equation}
Assuming the theory confines (due to an appropriate choice for the number of flavors) the baryon is a degree of freedom of the low energy effective field theory. Owing to Fermi statistics, the resulting operator is in the $N$-fold fully symmetrized representation:
\begin{equation}
\mathrm{Sym}^{N} \left( \frac{1}{2} \otimes \overline{N_f} \right),
\end{equation}
in the obvious notation. In particular, one can conceive of embedding $SU(2)_L$ in the diagonal $SU(N_f)$ flavor symmetry; adding additional interaction terms can also break $SU(N_f)_{\mathrm{diag}}$ down to $SU(2)_L$. The resulting bound state can thus be used to produce a field in the $N$-fold symmetrized product of $SU(2)_L$. Setting $N = 4$ would, for example, realize a $\mathbf{5}$-plet of $SU(2)_L$. This is not quite what we need, however, because the spacetime spin is also symmetrized, and would also result in a spin $N/2$ mode. It is also worth noting that the baryon constructed in this way is accompanied by many lighter degrees of freedom, including mesons in the adjoint representation of $SU(N_f)_{\mathrm{diag}}$.

The limitation of this warmup example is that we have introduced many fields which transform in a single big flavor symmetry. This in turn required us to full symmetrize all the flavor and spacetime spin indices simultaneously.

There are many ways to improve on the above situation to produce composite operators with high-dimensional flavor symmetry representations, but with low spacetime spin. As an example, one can consider composite operators built from both spin $1/2$ and spin $0$, as frequently occurs in supersymmetric gauge theories. For example, the scalar ``baryon operator'' of supersymmetric QCD is instead in the fully anti-symmetric representation of $SU(N_f)$.\footnote{This happens because due to spacetime spin-statistics now we need to anti-symmetrize all flavor indices.} This is not quite the representation we want, but it at least goes to the point that we can in principle build bigger representations.

To improve on these constructions, we can consider models in which the diagonal flavor symmetry is instead $SU(2)_1 \times ... \times SU(2)_K = SU(2)^K$ for some $K > 0$, viewing $SU(2)_L$ as the diagonal subgroup of $SU(2)^K$. Labelling each set of bifundamentals as $\psi^{(i)}$, we can again make composite states but now where the spacetime spin is not correlated with the representation for our flavor symmetry.

A general comment at this point is that in the process of building our desired high-dimension representation, we seem to inevitably encounter other composite states which are in lower-dimension representations of the
flavor symmetry. For example, in real world QCD, we have baryons such as the decuplet and octet of $SU(3)_{\mathrm{flavor}}$ but we also have the pions (as members of a different octet), which are significantly lighter. The general expectation, in fact, is that one ought to expect an entire tower of resonances of varying spins and representations.

Building examples with such high-dimensional representations can also be carried out
in string-based constructions.
For example, in many F-theory constructions where
the Standard Model is realized via a stack of intersecting / magnetized
7-branes, we can introduce an extra sector via a probe D3-brane. Examples of
strongly coupled extra sectors realized in this fashion include
\cite{Heckman:2010fh, Heckman:2011hu, Heckman:2015kqk, Heckman:2022the}. In
this setting, the extra sector will typically support many light degrees of
freedom in high-dimensional representations of the Standard Model gauge
group; this is in some sense a generic feature of having a conformal field
theory (CFT)\ with a flavor symmetry. There is a priori no reason that the higher-dimensional
representations will be accompanied by a high spacetime spin so there is generically no
issue with generating many high-dimensional representations.\footnote{A case of particular interest in 4D GUT Model building is the $\mathbf{126}$ of $SO(10)$, i.e., the five-index anti-symmetric representation subject to a self-duality constraint. It is tempting to identify the $\mathbf{126}$ with the ``baryon'' of an appropriate strongly coupled sector, but even this is more challenging to realize than one might have initially thought. For example, one might consider starting with an $Sp(N)$ gauge theory with $2F$ quarks in the fundamental representation. The issue is that this theory only has a mesonic branch; the baryonic branch is altogether absent. See reference \cite{Intriligator:1995ne} for further discussion of this theory.}

What then, is the basis of our claim that realizing $n$-plet scenarios is actually difficult / well nigh impossible in string theory? The main point is that in all of these examples the $n$-plet is never in isolation. Rather, it is simply part of a large tower of resonances in a strongly coupled sector that also includes lighter / comparable mass states in lower-dimensional representations. What we are really after, then, is a decoupling limit where \textit{all} of the other resonances other than the $n$-plet can be decoupled from the spectrum.

This is far more challenging to arrange. To illustrate some of the issues one confronts in trying to decouple
everything but an $n$-plet, it is helpful to keep in mind the QCD inequalities\footnote{See e.g.,
\cite{Weingarten:1983uj, Vafa:1983tf, Vafa:1984xg, Vafa:1984xh, Witten:1983ut,
Detmold:2014iha} and \cite{Nussinov:1999sx} for a review.} for a vectorlike gauge theory. In this case, positivity of the path integral measure in Euclidean signature leads to rather stringent constraints on the relative masses of the pions and baryons, and similar considerations are expected to hold for more ``exotic'' representations of the flavor symmetry of the sort of relevance to us. The main lesson in these theorems is that the pion ends up being lighter than the baryon.\footnote{There has been some work on developing a pionless effective field theory for nucleons \cite{Bedaque:2002mn, Hammer:2010kp, Griesshammer:2012we, Konig:2015aka}. In these models, however, one is in a very low energy non-relativistic regime. In the relativistic regime of the sort exclusively considered in this work, one still needs to include the pion as a mediator for nucleon interactions.} As such, the \textquotedblleft just $n$-plet\textquotedblright%
\ scenario is actually not realized by such vectorlike theories. A related comment is that insofar as the baryons of a QCD-like theory can be treated as solitonic excitations, namely skyrmions, they must necessarily be heavier than the pions of the theory.\footnote{We thank Z. Komargodski for a comment on this point.} It also runs counter to large $N$ counting arguments of the sort presented in \cite{Witten:1979kh}.

In this regard an interesting situation to consider is that of supersymmetric QCD with $N_f = N_c+1$ flavors. In this case, the existence of a baryonic and mesonic branch for the moduli space \cite{Seiberg:1994pq} means that at least in this limit, the baryons and mesons of SQCD are both exactly massless. In principle adding a non-supersymmetric mass term for the squarks could lift this degeneracy, and a priori one could imagine the baryon being lighter than the pion.\footnote{We thank T. Dumitrescu for this helpful comment.} See for example \cite{Aharony:1995zh, Arkani-Hamed:1998dti} for some discussion of such non-supersymmetric deformations. Even so, it seems rather implausible that in these scenarios one can fully decouple the other representations. Said differently, all of these states are still filling out a tower of states and cannot be decoupled from one another.

An additional comment here is that in the related examples discussed above
where we considered strongly coupled bound states of open strings (in IIB\ /
F-theory) as well as higher Kac-Moody level constructions,
the bottom up characterization of such states
inevitably appears to involve a compositeness scale of some kind. Of course,
in these latter examples the compositeness scale is all the way up at the
string scale, but from a bottom up perspective one can consider taking this
far lower, and this appears to also be realizable in string constructions (as
we have just shown).

Summarizing, while it is in principle possible to realize
high-dimensional representations in string constructions, we are not aware of any example which includes
the Standard Model as a subsector. Further, in approaches to realizing an $n$-pet of $SU(2)_L$ with $n \geq 5$,
we inevitably appear to encounter a whole tower of extra states including those in lower-dimensional
representations.\footnote{In the broader context of quantum gravity, there are
strong expectations based on \textquotedblleft completeness of the
spectrum\textquotedblright\ that there must be physical states in every
possible representation of a given gauge group \cite{Polchinski:2003bq, Banks:2010zn}.
For our present considerations, this is too weak a constraint since these extra states can in principle be very heavy with masses which are not tied to the specific type of
representation under consideration.}

This, of course, falls short of an actual no-go theorem, but all of the evidence points to this being an extremely challenging scenario to accommodate. Said differently, detection of the ``just $n$-plet scenario'' would
eliminate the entire known landscape of string vacua.

 \section{ATLAS Criteria for Disappearing Track Search} \label{app:ATLASTRACK}

In this Appendix we briefly list the ATLAS criteria for their disappearing track search \cite{ATLAS:2022rme}:

\begin{enumerate}
    \item Pseudo-rapidity in the range: $0.1 < \abs{\eta} < 1.9$
    \item Hits in 4 consecutive Pixel layers and none in the SCT
    \item $p_T > 20$ GeV
    \item Longitudinal impact parameter: $\abs{d_0 / \text{err} \qty(d_0)} < 1.5$
    \item Transverse impact parameter: $\abs{z_0^{\text{HS}}} \sin{\theta} < 0.5$ mm
    \item The deviation between the measured data and the tracklet fits lies within a likelihood of $10
\%$, given the tracklet. That is the chi-squared probability $\qty(\chi^2, \text{ndf}) > 0.1$.
    \item Isolation from other tracks in the event. The sum of standard track momenta in a cone of $\Delta R = 0.4$ about the disappearing track must be less than $4\%$ of the disappearing track momenta: $p_T^{\text{cone} 40} /p_T < 0.04$
    \item Isolation from calorimeter deposits. The sum of transverse energy in a cone of $\Delta R = 0.2$ about the disappearing track must be less than 5 GeV: $E_T^{\text{topoclus20}} < 5 \, \mathrm{GeV}$
    \item Overlap removal, discard event if an electron or muon is within $\Delta R = 0.4$ of the disappearing track
\end{enumerate}
In our own simulation we implement items 1, 3, 8, and 9.

\newpage

\bibliographystyle{utphys}
\bibliography{Strumion}

\end{document}